\documentclass[twoside,twocolumn,english,prl,aps]{revtex4}
\usepackage[T1]{fontenc}
\usepackage[latin1]{inputenc}
\usepackage{graphicx}

\makeatletter

\usepackage{babel}
\makeatother
\begin{document}

\title{Crossover in the Structure Between Bloch and Linear Domain Walls}

\author{Pavel Krotkov%
\footnote{E-mail: krotkov@magnet.fsu.edu%
}}

\affiliation{National High Magnetic Field Laboratory, Florida State University,
Tallahassee, Florida 32310, USA,}

\affiliation{L.D. Landau Institute for Theoretical Physics, Russian Academy of
Sciences, 2 Kosygina st., 117334 Moscow Russia}

\begin{abstract}
Near the Curie temperature of a ferromagnet the form of a domain wall
changes from the Bloch type to (asymptotically) the linear Zhirnov
wall. Unlike the simple 180$^{\circ}$ rotation of the magnetization
vector in a Bloch wall, its absolute value diminishes near the center
of the wall. This leads to a decrease of the total transverse component
of the exchange field inside the wall and to an increase of mistracking
of the spins of the electrons traversing the wall. This mechanism
may help explain large magnetoresistance of domain walls in thin nanowires,
as the Curie temperatures of low-dimensional nanostructures are known
to be lower than in bulk ferromagnets while the anisotropy energy
stays virtually unchanged.
\end{abstract}

\date{Jan 22, 2004}

\maketitle
\newcommand{\grad}{\beta}

\newcommand{\anis}{\kappa}

\newcommand{\f}[1]{\mbox {\boldmath\(#1\)}}

\newcommand{\ts}[1]{\textstyle #1}

\newcommand{\const}{\mathop{\mathrm{const}}\nolimits}

\newcommand{\wbl}{w_{\mathrm{Bl}}}

\newcommand{\wzh}{w_{\mathrm{Zh}}}

Strong current interest in magnetoresistance of domain walls in metallic
ferromagnets is motivated by possible applications in magnetoelectronics.
But at the moment there remains difficulty in reconciling experimental
and theoretical results. Different experimental groups reported both
positive and negative contribution of a single wall to resistance
(see, e.g., the review in \cite{kent01}). The theory of ``mistracking''
of the electron spin, e.g. the inability of its precession to track
the changing local exchange field as the electron traverses the wall,
could explain \cite{levy97} a small ($\sim2$\%) increase of the
wall resistance \cite{Gregg96}. However, various other contributions
to resistance of the same order of magnitude that are either negative
\cite{tatara97} or can have both signs \cite{vanGorkom99}, were
proposed, that could instead lead to a decrease in the domain wall
resistance \cite{Hong98,Ruediger98}. We want to show below that the
controversy of the experimental results may be resolved in terms of
the wall structure in different experiments. 

Magnetoresistance of domain walls in Ni nanocontacts was found to
be very large \cite{Garcia99,Chopra02}. The theory \cite{bruno99}
showed that geometrically constrained domain walls in atomic point
contacts become much sharper than in the bulk or thin films, with
the width on the scale of the size of the constriction, thus enhancing
the mistracking and, consequently, the resistance. 

A very large increase (100\%-600\%) in the resistivity of domain walls
was also observed in 35 nm Co nanowires \cite{Ebels00}. The authors
suggested that the mechanism behind this raise may be similar to the
giant magnetoresistance in current perpendicular to the plane geometry
(GMR-CPP). I.e., the mistracking of the passing electrons causes spin
accumulation at the domain walls which extends on the scale of the
spin diffusion length much larger than the domain wall width and gives
rise to the large resistance. In a theoretical study \cite{dzero03}
it was found that spin accumulation on Bloch domain walls is insufficient
to give rise to large magnetoresistance, and was suggested that the
domain wall may be of Zhirnov linear type \cite{zhirnov59}. 

In a linear domain wall magnetization remains along the easy axis
and is inverted by diminishing the absolute value and passing through
zero. The absence of a transverse component of the exchange field
in a linear wall eliminates the torque rotating the spin of the traversing
electron, and maximizes the mistracking. A linear wall is energetically
more favorable than the Bloch one in the bulk of a ferromagnet when
the magnetic energy becomes weaker than the anisotropy, e.g. near
the Curie temperature in bulk ferromagnets as in Zhirnov's original
paper \cite{zhirnov59}. Since the Curie temperature of thin films
and wires is known to be considerably lower than those of bulk ferromagnets,
magnetic energy can become comparable to or even less than the anisotropy
in thin nanowires.

In this paper we describe the transition between Bloch and linear
domain walls with the relative change of the coefficients of magnetic
energy and anisotropy in the Landau functional. We show that the magnetization
always rotates in a transition layer of the Bloch wall width, but
its absolute value could change. If the magnetic energy is comparable
to or less than the anisotropy, another scale, the Zhirnov linear
wall width, greater than the Bloch wall width, appears in the problem.
The absolute value of magnetization then diminishes from the values
in the domains toward the region where the inversion occurs on the
scale of Zhirnov linear wall width. So the transverse component of
the average magnetic moment in the wall diminishes, which leads to
increased electron spin mistracking and higher wall resistance.

Quite generally, equilibrium magnetic domain structure is determined
from the minimum of the total energy of a ferromagnet below the Curie
temperature including exchange, anisotropy, magnetostatic (stray,
or dipole-dipole) and magnetoelastic energies. In the simplest case
of uniaxial crystals average magnetic moment along the anisotropy
axis (easy-axis) $\hat{\mathbf{z}}$ is oriented oppositely in the
neighboring domains. This case corresponds to a positive anisotropy
coefficient $\anis>0$ in the Landau expansion of the density of ferromagnetic
free energy:\begin{equation}
\mathcal{F}=\mathcal{F}_{0}+A\mathbf{M}^{2}+B\mathbf{M}^{4}+\ts\frac{1}{2}\grad\left(\partial_{i}\mathbf{M}\right)^{2}+\ts\frac{1}{2}\anis\mathbf{M}_{\perp}^{2}.\label{eq:F}\end{equation}
Here $\mathbf{M}_{\perp}=\mathbf{M}-M_{z}\hat{\mathbf{z}}$ is the
in-plane component of the average magnetic moment $\mathbf{M}$. For
positive $\anis$ the minimums $\mathcal{F}_{\const}=\mathcal{F}_{0}-|A|M_{s}^{2}/2$
of (\ref{eq:F}) among the spatially uniform solutions are reached
when \begin{equation}
\mathbf{M}=\pm M_{s}\hat{\mathbf{z}}\qquad\mathrm{with}\qquad M_{s}^{2}=|A|/2B\label{eq:Muni}\end{equation}
(in a ferromagnetic state $A<0$). 

Consider a transition layer (a domain wall) between two regions with
uniform equilibrium magnetization (\ref{eq:Muni}) in the bulk of
a ferromagnet. In this case \textbf{$\mathbf{M}$} is varying only
in the direction $\hat{\mathbf{l}}$ perpendicular to the plane of
the wall. If we introduce the normalized magnetization $m=M/M_{s}$
and the polar coordinates $(\theta,\varphi)$ of $\mathbf{M}$ with
respect to $\hat{\mathbf{z}}$, free energy (\ref{eq:F}) is rewritten
as\begin{eqnarray}
\mathcal{F} & = & \mathcal{F}_{\const}+\ts\frac{1}{2}M_{s}^{2}\bigl[|A|(1-m^{2})^{2}\nonumber \\
 &  & +\grad(\dot{m}^{2}+m^{2}\dot{\theta}^{2})+(\anis+\grad\dot{\varphi}^{2})m^{2}\sin^{2}\theta\bigr],\end{eqnarray}
where a dot over a symbol implies a derivative over $l$. An equilibrium
wall structure $m(l)$, $\theta(l)$, and $\varphi(l)$ has to be
found from this functional variationally. 

The Euler-Lagrange equation $\delta\mathcal{F}/\delta\varphi=0$ gives
$\ddot{\varphi}=0$, hence $\dot{\varphi}=\const$. A minimum of $\mathcal{F}$
is obtained when $\dot{\varphi}\equiv0$, i.e., the magnetization
stays in a plane, which we will choose to be the $xz$-plane, so that
$\varphi\equiv0$.

Equation $\delta\mathcal{F}/\delta\theta=0$ gives $\grad\ddot{\theta}=\anis\sin\theta\cos\theta$.
Its first integral \begin{equation}
I=\grad\dot{\theta}^{2}-\anis\sin^{2}\theta=\const\end{equation}
 vanishes in the domains at $\pm\infty$, therefore $I\equiv0$. Whence
we find that the solution with boundary conditions $\cos\theta(\pm\infty)=\pm1$
and the center of the wall at the origin is \cite{landau35}\begin{equation}
\cos\theta=\tanh(l/w_{\mathrm{Bl}}).\label{eq:thetaSol}\end{equation}
The spatial scale $\wbl=\sqrt{\grad/\anis}$ is the width of the Bloch
wall. It is determined by a competition between the inhomogeneous
exchange interaction which tends to increase $\wbl$ and of the magnetic
anisotropy which decreases $\wbl$. The direction of magnetization
rotates from $0$ to $\pi$ on the scale of $\wbl$.

We still have to find the normalized absolute value $m$ of the magnetization
from $\delta\mathcal{F}/\delta m=0$: \begin{equation}
\grad\ddot{m}=m(\grad\dot{\theta}^{2}+\anis\sin^{2}\theta)-2|A|m(1-m^{2}).\end{equation}
Substituting (\ref{eq:thetaSol}) we arrive at \begin{equation}
\grad\ddot{m}=2m\left[\frac{\anis}{\cosh^{2}(l/\wbl)}-|A|(1-m^{2})\right].\label{eq:m}\end{equation}
The variation of $m$ is determined by a competition of the two terms.
The first is the anisotropy which acts only in the region $l\lesssim\wbl$
near the center of the wall, where magnetization deviates from the
easy axis. The second term comes from the change in the absolute value
of the magnetization. It acts on a scale of the width of Zhirnov wall
$\wzh=\sqrt{\grad/|A|}$ which may be smaller or greater than $\wbl$
depending on the parameter $|A|/\anis$ which gives the relative strength
of the magnetic energy and anisotropy. 

Deep below the Curie temperature one can neglect the variation of
the absolute value of the magnetization $m$. In this case the solution,
calculated by Landau and Lifshitz \cite{landau35}, is the Bloch wall
in which $m$ remains constant and the angle $\theta$ inverts on
the scale of $\wbl$. Indeed, when $|A|\gg\anis$, the first term
in (\ref{eq:m}) may be neglected and the equation\begin{equation}
\grad\ddot{m}=-2|A|m(1-m^{2})\label{eq:slowm}\end{equation}
 has only the trivial solution $m\equiv1$ satisfying the boundary
conditions $m(\pm\infty)=1$. 

\begin{figure}
\includegraphics[%
  width=1.0\columnwidth]{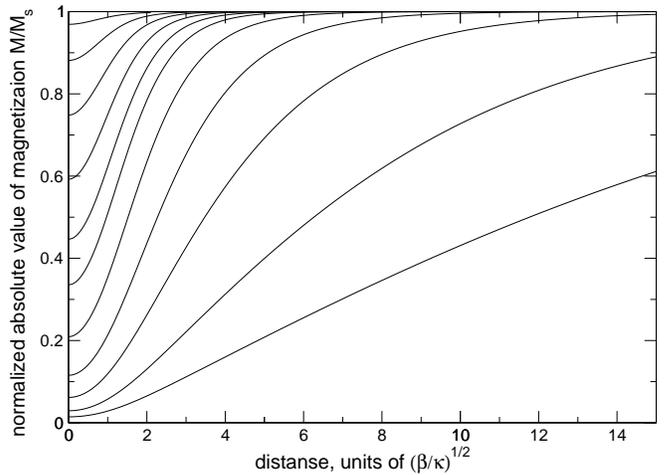}

\caption{The variation of the absolute value of magnetization through domain
walls in a bulk ferromagnetic. The parameter $\wzh/\wbl$ is 0.25,
0.5, 0.75, 1, 1.25, 1.5, 2, 3, 5, 10, and 20 for the curves from top
to bottom. \label{cap:The-variation-of}}
\end{figure}

Zhirnov \cite{zhirnov59} considered the domain wall structure near
the Curie temperature, when $A=\alpha(T-T_{c})$ is smaller than the
anisotropy, and it becomes energetically more favorable to diminish
$m$ rather than to tilt magnetization from the easy axis. Zhirnov
linear wall is most easily obtained if we omit the first term in (\ref{eq:m})
altogether and allow $m$ to change sign. To pass to the limit of
Zhirnov wall in the solution of (\ref{eq:m}) with the chosen parametrization
of $\mathbf{M}$, although possible, is not straightforward, and we
postpone the discussion of this academic problem until the end of
the paper.

For arbitrary ratios $\anis/|A|$ the solution for the absolute value
of magnetization can only be obtained numerically. The curves $m(l)$
for several values of the parameter $\wzh/\wbl=\sqrt{\kappa/|A|}$
are plotted in Fig. \ref{cap:The-variation-of}. The increase in the
relative value of the anisotropy $\anis/|A|$ describes the transition
from the Bloch to linear wall. 

We see that for small $\wzh/\wbl$, when the anisotropy is relatively
weak, the first term in (\ref{eq:m}) only causes a small indentation
in $m$ in the region $l\lesssim\wbl$ near the wall center where
the magnetization inverts its direction. A measure of the depth of
this indentation is the absolute value of the magnetization in the
center of the wall $m(l=0)$, which is always $<1$ for finite anisotropy
$\anis/|A|$.

As $\wzh$ approaches $\wbl$, $m(l=0)$ goes down. Nevertheless,
the width of the indentation remains almost constant and equal to
$\wbl$. At approximately the point when $\wzh/\wbl\approx2\div3$
the second term in (\ref{eq:m}) takes over the variation of $m$.
From this point on, the width of the indentation grows approximately
linearly with $\wzh$ and $m(l=0)$ continues to fall slowly, approaching
zero asymptotically as the wall goes to a pure linear limit $\anis\gg|A|$.

\begin{figure}
\includegraphics[%
  width=1.0\columnwidth]{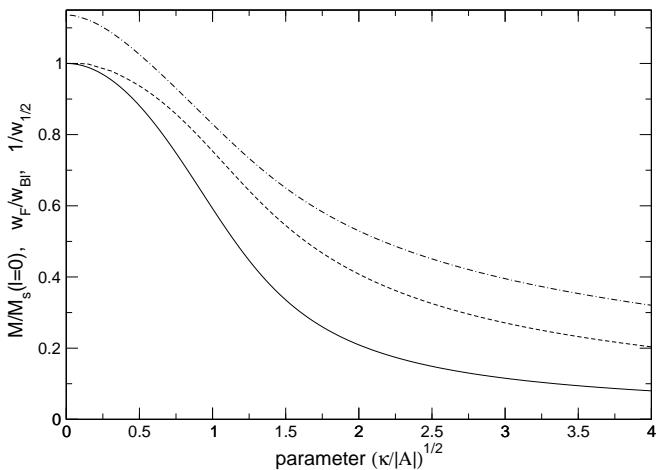}

\caption{Normalized absolute value of magnetization at the center of the wall
$m(l=0)$ (solid line), the flux width $w_{F}$ (\ref{eq:wF}) of
the wall (dashed line), and the inverse half-width of the wall (dash-dotted
line) as functions of the parameter $\wzh/\wbl=\sqrt{\kappa/|A|}$.\label{cap:Normalized-absolute-value}}
\end{figure}

Quantitatively, this transition is illustrated in Fig. \ref{cap:Normalized-absolute-value},
where the absolute value of the magnetization at the center of the
wall $m(l=0)$ (solid line) and the inverse half-width of the indentation
(dash-dotted line) are plotted vs. $\wzh/\wbl$.

There are clearly two spatial scales in the problem: angle $\theta$
is inverted in a layer of the width $\wbl$; also, if $\wzh>\wbl$,
the absolute value of magnetization changes with the characteristic
length of $\wzh$. Which of the scales defines the wall width depends
on the context, and various definitions were proposed in the literature
\cite{hubert98}. All of them, however, are based on the variation
of the angle $\theta(l)$ only and are not particularly suitable for
the wall in question. The definition of the wall width based on the
total wall flux \begin{equation}
w_{F}=\frac{1}{\pi}\int_{-\infty}^{\infty}m(l)\sin\theta(l)dl\label{eq:wF}\end{equation}
seems a viable suggestion. It describes contrast in Bitter pattern
experiments \cite{hubert98}. And, since (\ref{eq:wF}) gives the
integral of the transverse component of magnetization in a wall, it
serves as a qualitative measure of mistracking, and, hence, of the
wall resistance.

The flux width of the wall is plotted in units of $\wbl$ as dashed
line in Fig. \ref{cap:Normalized-absolute-value}. It changes from
$\wbl$ for the limiting case of a pure Bloch domain wall to zero
for a pure Zhirnov linear wall. Thus with the increase of the anisotropy,
as the wall transforms into a linear one, the transverse component
of magnetization in the wall diminishes, so the electron spin becomes
less able to track the changing local exchange field when it traverses
the wall, and the wall resistance increases. 

In a finite sample the structure of a domain wall is found from a
minimum of the sum of the local free energy (\ref{eq:F}) and of the
non-local dipolar energy depending on the form of the sample. This
is a highly non-trivial task even for the simple geometry of cylindrical
wires with the easy axis along the wires used in \cite{Ebels00}.
A rough estimate of the importance of the dipolar energy is given
by $2\pi M_{s}^{2}$ multiplied by the wall width compared to the
surface tension of the wall. For a Bloch wall this leads to the usual
``quality factor'' criterion: if $2\pi M_{s}^{2}/\anis$ is less than
unity, the dipolar interaction can be neglected, otherwise not. For
a Zhirnov wall the ratio of the dipolar energy of the wall to its
tension calculated by neglecting the dipolar contribution is $2\pi M_{s}^{2}/|A|$.
Since in the Zhirnov regime $|A|\ll\anis$, this is a stricter requirement.
So, we conclude, the magnetostatic energy does play a role in the
form of the domain wall in finite samples in the regime close to the
pure linear Zhirnov wall. However, this is a problem of the next level
of complexity that has to be studied separately.

To conclude, in an equilibrium domain wall in a bulk ferromagnet the
magnetization is flipped on the scale of the Bloch wall width. If
the anisotropy is comparable or stronger than the magnetic energy
the absolute value of magnetization in the region where it is flipped
is less than the value in the domains. The decrease in the absolute
value occurs on the scale of the width of Zhirnov linear wall. This
decrease diminishes the torque acting on the spin of an electron traversing
the wall, impairs the ability of the electron spin to track the changing
magnetization, and leads to a greater spin accumulation GMR effect,
and thus to a greater wall resistance. This may contribute to the
observed large magnetoresistance of domain walls in 35 nm Co nanowires
\cite{Ebels00}. A quantitative estimate of the effect is hindered
by the absence of data on the Curie temperature of nanowires. Encouraging,
though, is the absence of large magnetoresistance in slightly thicker
wires of 50 nm in diameter \cite{Ebels00} which presumably have greater
$T_{c}$. 

In the end of the paper, we show how a formal passing to the limit
of pure Zhirnov wall may be done in the solution of (\ref{eq:m}).
Zhirnov linear wall is realized in the limit $\wzh>\wbl$. Then outside
of the region of the inversion of $\theta$, $|l|\lesssim\wbl$, the
slow variation of $m$ is still described by (\ref{eq:slowm}). The
non-trivial solutions with the boundary conditions $m(\pm\infty)=1$
are respectively \begin{equation}
m(l)=\tanh\frac{l_{0}\pm l}{\wzh}.\label{eq:twoSolutions}\end{equation}
The first term of (\ref{eq:m}) is non-zero only closer than $\wbl$
to the center of the wall. To describe the variation of $m$ on the
scale of $\wzh$ it may be substituted by a delta-function: $\cosh^{-2}(l/\wbl)\to2\delta(l/\wbl)$.
Thus the two solutions (\ref{eq:twoSolutions}) need to be matched
at the origin so that the derivative $\dot{m}(0)$ had a jump \begin{equation}
\dot{m}(+0)-\dot{m}(-0)=4m(0).\label{eq:11}\end{equation}
Since for small $l_{0}$ Eq. (\ref{eq:twoSolutions}) gives $m(0)\approx l_{0}/\wzh$
and $\dot{m}(0)\approx\pm1/\wzh$, condition (\ref{eq:11}) corresponds
to a choice of $l_{0}=\wbl/2$. So the solution is given by\begin{equation}
m(l)=\tanh\frac{\wbl/2\pm|l|}{\wzh}.\end{equation}
 On the scale of $\wzh$ one may neglect $\wbl/2$ compared to $|l|$,
and we finally have\begin{equation}
m(l)=\tanh|l|/\wzh.\end{equation}
The law of inversion (\ref{eq:thetaSol}) of the angle $\theta$ on
the scale $\wzh$ is\begin{equation}
\theta(l)=\pi\Theta(l),\end{equation}
where $\Theta(x)$ is the Heaviside function. In a pure Zhirnov wall
magnetization never leaves the easy axis and only changes its direction
passing through zero. Hence the chosen parametrization of $\mathbf{M}$
by $m$, $\theta$, and $\varphi$ is not very convenient for such
a transition. That's why an indirect procedure above was needed.

We gratefully acknowledge discussions with L. P. Gor'kov. Work was
supported from DARPA through the Naval Research Laboratory Grant No.
N00173-00-1-6005.

\end{document}